\documentclass[prl,showpacs]{revtex4}
\usepackage{amsmath}
\usepackage{graphicx}
\usepackage{epsfig}
\usepackage{dcolumn}
\usepackage{bm}
\usepackage{dcolumn}
\usepackage{amsfonts}
\usepackage{amssymb}

\begin{document}
\title{Spontaneous symmetry breaking approach to La$_2$CuO$_4$ properties:\\ hints
for matching the Mott and Slater pictures}
\author{Alejandro Cabo-Bizet$^*$ and  Alejandro Cabo-Montes de Oca$^{**}$}

\affiliation{ $^*$Departamento de F\'{\i}sica,  Centro
de Aplicaciones Tecnol\'{o}gicas y Desarrollo Nuclear (CEADEN),
Calle 30, esq. a 5ta, La Habana, Cuba.}

\affiliation{$^{**}$Grupo de F\'isica Te\'orica, Instituto de
Cibern\'etica Matem\'atematica y F\'{i}sica (ICIMAF),\\ Calle E, No.
309, entre 13 y 15, Vedado, La Habana, Cuba.}


\begin{abstract}
 Special solutions of the Hartree-Fock (HF) problem  for Coulomb
interacting electrons, being described by a simple model of the Cu-O planes in
La$_{2}$CuO$_{4} $,  are presented. One of the  mean field states obtained, is  able to predict  some of the basic properties of this material, such as its insulator character and the antiferromagnetic order. The natural appearance of pseudogaps in some states of this compound  is also indicated by
another of the HF states obtained. These surprising results follow after eliminating spin and crystal   symmetry restrictions which are usually imposed on the single particle HF orbitals, by means of  employing a  rotational invariant formulation of the HF scheme which was originally introduced  by Dirac. Therefore, it is exemplified  how, up to now being considered strong correlation effects,  can be described by improving the HF solution of the physical systems.  In other words, defining the correlation effects as such ones shown by the physical system and which are not predicted by the best HF (lowest energy) solution,  allows to explain  currently  assumed as strong correlation properties,  as simple mean field ones. The discussion also helps to clarify the role of the antiferromagnetism and pseudogaps in the physics  of the HTSC materials and indicates a promising way to start conciliating the Mott and Slater pictures for the description of the transition metal oxides and other strongly correlated electron systems.

\end{abstract}

\maketitle
The Hubbard  models in the theory of strongly correlated
electron systems are notably successful
\cite{mott,slater1,peierls,hubbard,anderson,kohn,
imada,dagoto,yanase,damascelli}. They remarkable reproduce the
properties of Mott insulators, such as metal-transitions oxides and
copper-oxygen layered HTc compounds \cite{imada}. However, the
search for approaches having more basic foundations do not cease,
hoping that they could open the way for obtaining more exact
and specific results \cite{terakura}. The so called  \textit{first
principle or Slater} procedures,  are electronic structure calculations that
begin by considering  the interactions among electrons or atoms in vacuum. The
study of the band structure they predict, in principle should offer
a way for the precise determination of the
physical properties of each material \cite{matheiss,terakura,kohn1}.
However, the above mentioned potentialities had been failing in
describing many systems showing
strong correlation effects \cite{imada}. \\
\indent The motivation of the present letter emerged from a
suspicion that the Hartree Fock (HF) method, could had been
underestimated in its possibilities for helping in clarifying the
above mentioned difficulties \cite{dirac,slater1}. As a net result
of this work, we came to believe that a large deal of correlation
effects, can be described in the framework of the HF scheme, after
removing certain symmetry restrictions which obstacle the finding of
the best HF solutions. By example, it has been early shown by Slater
in Ref. \cite{slater1}, that sometimes the HF potential breaks the
symmetry of the original crystalline lattice, creating magnetic
structures and  gaps. This symmetry breaking effect has been also
more recently underlined and deepened in Ref. \cite{const1}.
However, the removal of the lattice symmetry restrictions alone had
not been able to describe the insulator properties of a large class
of materials
 \cite{mott,terakura,imada}. One of the central results of the
present investigation, is the identification of another kind of
symmetry restriction that seemingly had been overlooked for a long
time. It corresponds to the usual assumption about the necessary
$\alpha$ ($S_z=1/2$) or $\beta$ ($S_z=-1/2$) orientations of the
spin of each solution for the HF orbitals \cite{szabo,matheiss}. For
supporting this main statement, the work  considered the HF problem
as applied to a simple one band model of the superconducting
material La$_{2}$CuO$_{4}$\cite{pickett}, looking from the start for
single particle orbitals being non separable in their spacial and
spinor dependence. That is,  they will have the structure
$\phi(x,s)\neq\phi(x)\psi(s)$, i.e. the orbitals have not an
absolute common quantization direction for their electron spin.
Note, that to proceed in this way means not other thing that to
apply the Dirac's  formulation of the HF procedure \cite{dirac}. The
results, as it will be seen, are able to dscribe the basic
properties of the La$_{2}$CuO$_{4}$  \cite{pickett}.
  We would like to stress that the obtained results do not look so
radical if we consider the following circumstance: the correlation effects
are associated with the difference shown by the system's properties with
the ones predicted by the HF procedure applied to it.  Therefore, what we have argued here,
is simply that the there exist unexplored improvements of the mean field schemes which allow to
predict properties which  are currently considered as strong correlation
properties. These modifications of the HF method are related with usual constraint
which are imposed on the crystalline and spin properties of the single particle orbitals.
The possibility of their employment for solving  the longstanding debate between the Slater and Mott
pictures in the theory of transition metal oxides and HTc superconducting materials will be considered
elsewhere.

\begin{figure}[h]
\vspace{.1cm}
\par
\begin{center}
\includegraphics[width=1.3in,height=1.3in]{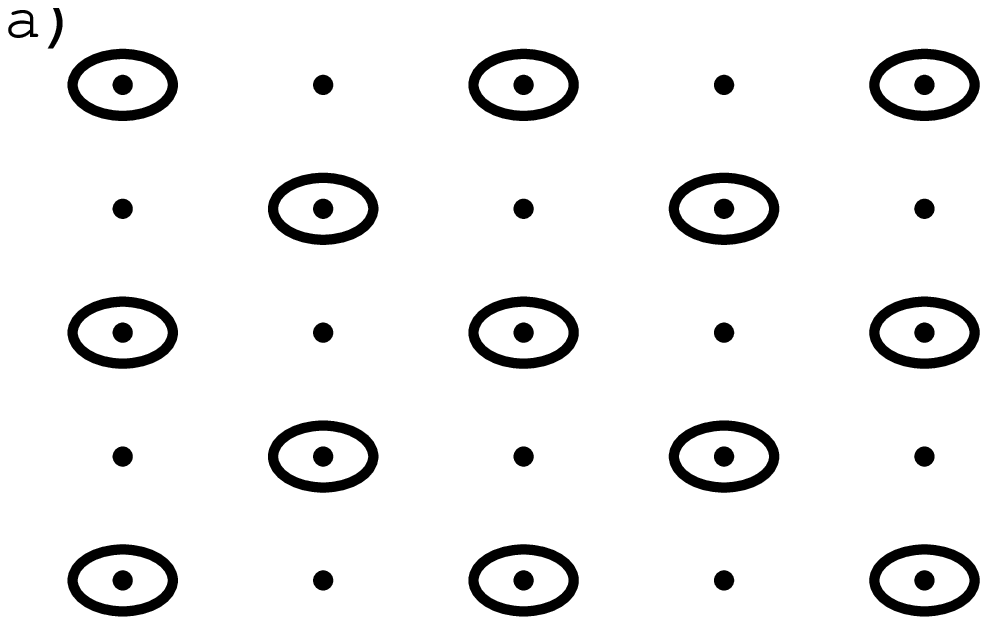}
\ \ \ \ \ \ \ \includegraphics[width=1.3in,height=1.3in]{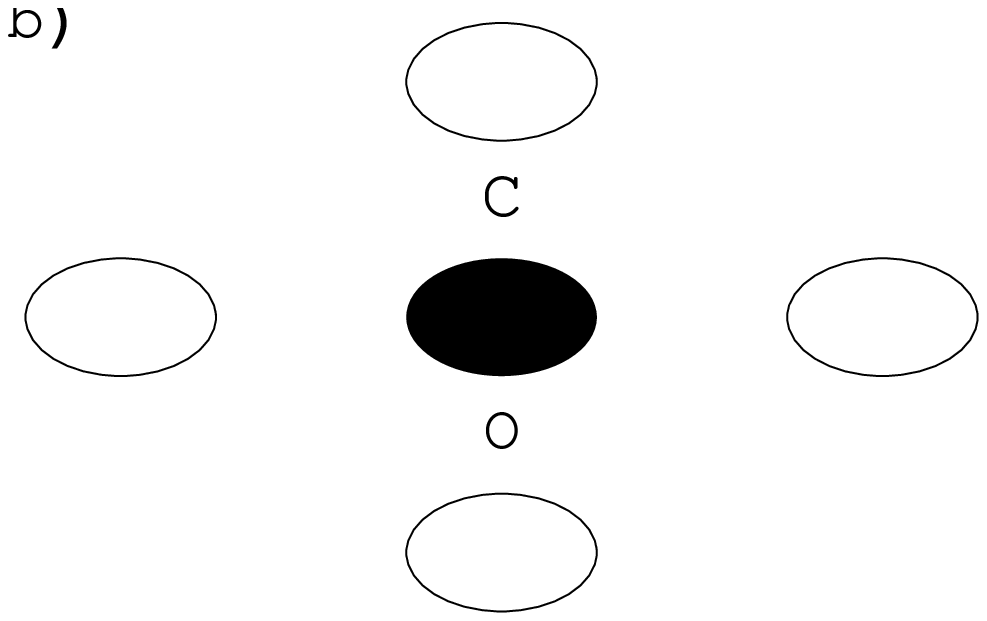}
\end{center}
\par
\vspace{.1cm}\caption{The figures show: a) The point lattice
associated to the Cu-O planes.  For removing the symmetry
restrictions, it will be helpful to separate the lattice in the two
represented sublattices; and b)
shows the corresponding base of the Cu-O planes.}%
\label{f:bravais}%
\end{figure}

In this letter, we will firstly describe the physical basis of the model employed
and after that, the results obtained for the various HF solutions
will be presented. For the details of the concrete solutions  we
refer to the extended version of this work in Ref.
\cite{sversion}. \\
\indent Let $\hat{\mathcal{H}}(x_{1},...x_{N})=\sum_{i}\hat{\mathcal{H}}_{0}(x_{i}%
)+\frac{1}{2}\sum_{j\neq i}V(x_{i},x_{j})$, be the N-electrons
system hamiltonian, including kinetic plus interaction with the
environment hamiltonian $\hat{\mathcal{H}}_{0}$, besides Coulomb
interaction among pairs of electrons $V$. The HF equations of motion
were solved by allowing for nonseparable solutions for the orbitals
$\phi_{\eta}(x,s)$, where $\eta=k_{1},...,k_{N}$ are  the labels of
the HF basis they form. \ The system of  HF equations is rotational
invariant when formulated in  the general HF procedure firstly
introduced by Dirac in Ref. \cite{dirac}. Its explicit expression
for the here considered problem can be found in Ref.
\cite{dirac,sversion}.  \\
\indent Let us first present the effective band model used to
describe the dynamic of the less bounded electrons in
La$_{2}$CuO$_{4}$. It is known that at low temperature
La$_{2}$CuO$_{4}$ is an antiferromagnetic-insulator \cite{pickett}.
However, in evident contradiction with the experiments the Linear
Augmented Plane Waves (LAPW) band calculations  predicts for it
metal and paramagnetic zero temperature properties \cite{matheiss}.
Nevertheless, such band studies results at least show that the
conduction electrons are strongly coupled to the Bravais lattice
centers of the copper oxygen planes. Clearly this tight-binding
behavior is determined by the interaction of the electrons with its
surrounding effective environment.  \ The less bounded electron in
the La$_{2}$CuO$_{4}$ molecule is the Cu$^{2+}$'s not paired one. At
difference from O$^{2-}$ ions, Cu ones do not have their last shell
(3d) closed. Those copper 3d electrons partially fill the last band
of La$_{2}$CuO$_{4}$ solid and in what follows they shall be
referred as: the electron gas. It seems appropriate to consider
those electrons as strongly linked to CuO$_{2}$ cells and with
special preference for the Cu centers \cite{anderson}. Thus, our
Bravais lattice is going to be the squared net coincident with the
array of copper sites (see the figure \ref{f:bravais}). Further, the
presence of electrons pertaining to the various fully filled bands
in the material plus the nuclear charges, will play a double role in
the model. Firstly, they will act as an effective polarizable
environment screening the field of the  charges in  the electron
gas. It will be  reflected by a dielectric constant $\epsilon$
reducing the Coulomb interaction. Secondly, the mean field created
by the environment will act as a periodic potential W$_{\gamma}$,
tight-binding the electrons to the Cu centers. The interaction
F$_{b}$ among the electron gas and a ''jellium'' neutralizing its
charges is also considered. It was be modeled as a gaussian
distribution of positive charges
$\rho_{b}(\mathbf{y})=\frac{1}{\pi b^{2}}\exp(-\frac{\mathbf{y}^{2}%
}{b^{2}}),$ surrounding each lattice point  and having a
characteristic radius \emph{b }. \ In resume, the free hamiltonian
of the model takes the form
\begin{align}
\hat{\mathcal{H}}_{0}(\mathbf{x})  & =\sum_{i=1}^{N}\frac{\hat{\mathbf{p}}%
_{i}^{2}}{2m}+W_{\gamma}(\mathbf{x})+F_{b}(\mathbf{x}),\label{freehamiltonian}%
\\
W_{\gamma}(\mathbf{x})  & =W_{\gamma}(\mathbf{x}+\mathbf{R}),\nonumber\\
F_{b}(\mathbf{x})  & =\frac{e^{2}}{4\pi\epsilon\epsilon_{0}}\sum_{\mathbf{R}%
}\int d^{2}y\frac{\rho_{b}(\mathbf{y}-\mathbf{R})}{|\mathbf{x}-\mathbf{y}%
|},\ b\ll p,\nonumber
\end{align}
where $\hat{\mathbf{p}}_{i}^{2}$ is the i-th electron's squared
momentum operator; \emph{m} is the electron mass; $\epsilon_{0}$ is
the vacuum permitivity and lattice vectors $\ \mathbf{R}=n_{x_{1}}
p\, \hat{\mathbf{e}}_{x_{1}}+$ $n_{x_{2}} p\, \hat{\mathbf{e}}_{x_{2}}$ with $n_{x_{1}}$ and $n_{x_{2}%
}$ being integers, move on Bravais lattice. It will be referred in
what follows as  the $\textit{the absolute}$ lattice. The versors $\hat{\mathbf{e}%
}_{x_{1}}$ and $\hat{\mathbf{e}}_{x_{2}}$ are resting on the
direction defined by the lattice's nearest neighbors (see figure
\ref{f:bravais} a)). The distance  between Cu nearest neighbours is
\ $p\approx\ $ 3.8 \AA  \, \cite{pickett}. The interaction among
pairs of electrons is taken in the form
$\ V(\mathbf{x},\mathbf{y})=\frac{e^{2}}{4\pi\epsilon\epsilon_{0}%
}\frac{1}{|\mathbf{x}-\mathbf{y}|},$ which includes the dielectric
constant associated to the presence of the effective environment. \\
\indent We are seeking here for HF solutions with orbitals having a
non separable spin and
 orbit structures. Thus, it was considered that the spin can show a
 different projection  for the different Wannier orbitals to be superposed
 for defining those orbitals. The spin for
each of them will be either or $\alpha$ or $\beta$ type, according
they are linked either to one or  the other of the two sublattices
shown  in figure \ref{f:bravais} a). The points of
these sublattices  were defined as follows: $\mathbf{R}^{(r)}=\sqrt{2}%
n_{1}p\ \hat{\mathbf{q}}_{1}+\sqrt{2}n_{2}p\ \hat{\mathbf{q}}_{2}%
+\mathbf{q}^{(r)}$, r = 1, 2; with $n_{1}$ and $n_{2}\
\in\mathbb{Z}$ , and \ in which
the vector $\ \mathbf{q}^{(r)}=0$ if $r=1$ and $\mathbf{q}^{(r)}%
=p\ \hat{\mathbf{e}}_{x_{1}}$ when $\ r=2;$ and  where $\hat{\mathbf{q}}%
_{1}$ and $\hat{\mathbf{q}}_{2}$ form the basis versors on each one
of them.
The searched solutions were constructed as eigenfunctions of the
operators $\hat{T}_{\mathbf{R}^{(r)}}$ belonging to the
 translation group transforming each sublattice on itself: $\hat{T}_{\mathbf{R}%
^{(r)}}\phi_{\mathbf{k},l}=\exp(i\ \mathbf{k}\cdot\mathbf{R}^{(r)}%
)\phi_{\mathbf{k},l}.$ Therefore we  imposed periodic boundary
conditions on the $\phi_{\mathbf{k},l}$ in the absolute lattice's
boundaries x$_{1}$= - p L and p L, L , x$_{2}$= -p L  and p L  (see
figure \ref{f:bravais} a)). This condition
determines the allowed set of \ quasimomenta $\mathbf{k=}\frac{2\pi}%
{Lp}\
(n_{1}\hat{\mathbf{e}}_{x_{1}}+n_{2}\hat{\mathbf{e}}_{x_{2}}),$with
$n_{1},\ n_{2}\in\mathbb{Z}$ and $-\frac{L}{2}\leq n_{1}\pm n_{2}<\frac{L}%
{2}.$ \ Note that we are demanding less than the full crystal
symmetry on the single particle states which we are looking for. Let
the single particle states represented in the explicitly non
separable form
\[
\phi_{\mathbf{k},\ l}(\mathbf{x},s)=\sum_{r,\sigma_{z}}B_{r,\sigma_{z}%
}^{\mathbf{k},\
l}\varphi_{\mathbf{k}}^{(r,\sigma_{z})}(\mathbf{x},s),
\]
where \emph{l} is the additional label needed for indexing the
stationary states. The tight binding Bloch basis
$\varphi_{\mathbf{k}}^{(r,\sigma_{z})}$ appearing  is  defined as
\begin{align}
\varphi_{\mathbf{k}}^{(r,\sigma_{z})}(\mathbf{x},s)  & =\sqrt{\frac{2}{N}%
}\ u^{\sigma_{z}}(s)\sum_{\mathbf{R}^{(r)}}\exp(i\ \mathbf{k}\cdot
\mathbf{R}^{(r)})\ \varphi_{\mathbf{R}^{(r)}}(\mathbf{x}),\nonumber\\
\hat{\sigma}_{z}u^{\sigma_{z}}  & =\sigma_{z}\ u^{\sigma_{z}},\nonumber\\
\varphi_{\mathbf{R}^{(r)}}(\mathbf{x})  & =\frac{1}{\sqrt{\pi a^{2}}}%
\exp(-\frac{(\mathbf{x}-\mathbf{R}^{(r)})^{2}}{2\ a^{2}}),\ a\ll
p,\nonumber
\end{align} where $N$ is the number of electrons in the electron gas,
$\hat{\sigma}_{z}$ is the spin z projection operator, where z is the
orthogonal direction to the copper oxygen CuO$_{2}$ planes;
$\sigma_{z}=$ -1, 1, are the eigenvalue of the previously mentioned
operator and \emph{r} = 1, 2, is the label which indicates each one
of the sublattices. As we are working at  half filling condition,
then $N$ is equal to the number of cells in the periodicity region
N$_{c}$. Note that int his simple model, the electron were considered as moving in two dimensions
and that gaussian T.B. Wannier orbitals had been assumed \cite{sversion}.
\begin{figure}[h]
\begin{center}
\includegraphics[width=1.3 in,height=1.3in]{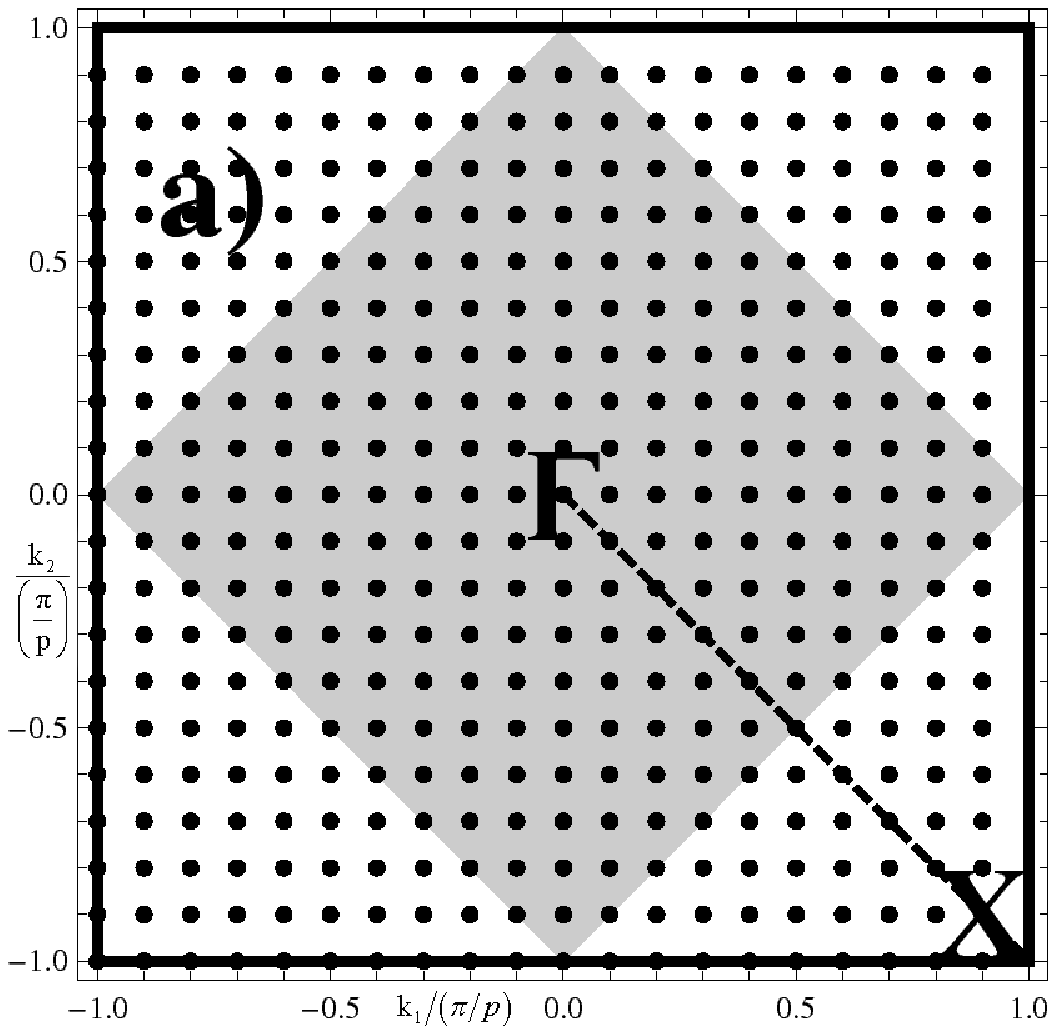}
\ \
\includegraphics[width=1.5in,height=1.5in]{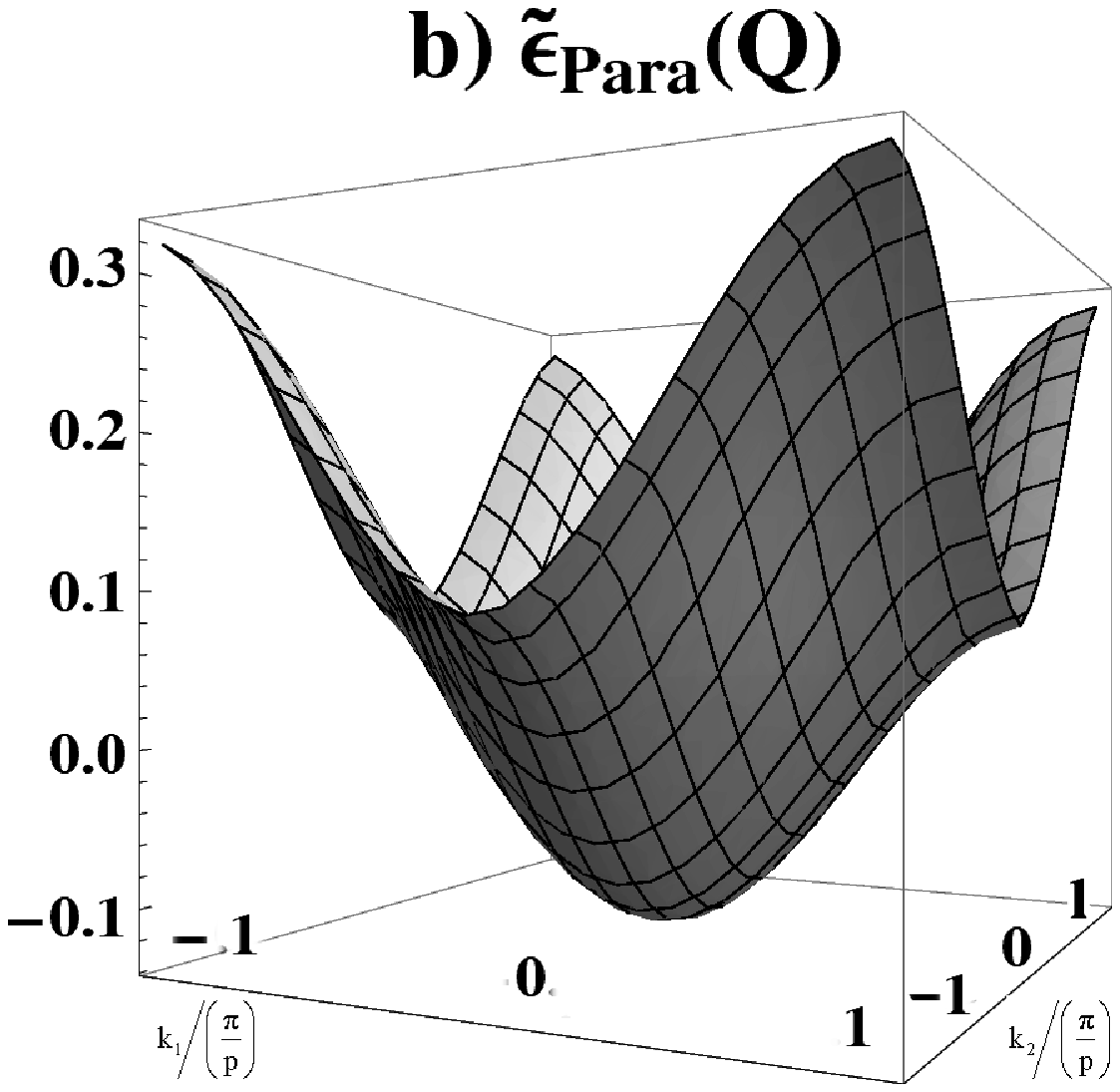}
\end{center}
\caption{ The figure 1 a) shows the Brillouin zone associated
$\textit{the absolute}$ lattice. The grey region signals the
Brillouin zone (B.Z.)of the sublattices. The unit of quasimommentum
is $\frac{\pi}{p}$. Figure b) shows the doubly degenerated bands
associated to a paramagnetic and metallic state. The zero energy
level in all the band diagrams is the Fermi energy
of the isolator and antiferromagnetic solution. }%
\label{f:bM}%
\end{figure} \\
\indent  The first solution considered, to be called the PM one, was
one having their orbitals  being eigenfunction of the maximal group
of translations leaving invariant the \textit{absolute} lattice.
That is, the maximum possible crystal symmetry is demanded. Also,
the spin structure of the  orbitals were assumed to be of $\alpha$
or $\beta$ \ types. For this case, figure \ref{f:bM} shows the
paramagnetic, metallic and doubly degenerate band obtained from an
iterative process of solving the HF equation in  Ref.
\cite{sversion}. For the case of $N=20\times20$ electrons, the
momenta of occupied states are shown in figure \ref{f:bM} a). They
lay inside the shadowed square of side $\sqrt{2}\pi/p$: the
Brillouin zone (B.Z) of the sublattices. The condition of
reproducing the bandwidth of 3.8 eV for the conduction band reported
in Ref. \cite{matheiss}, allowed to fix the dimensionless values of
the free parameters of our model to be:
$\epsilon$=10, $\widetilde{a}%
$=0.25, $\widetilde{b}$=0.05 and $\widetilde{\gamma}$=-0.03. See
Ref. \cite{sversion} for those dimensionless definitions.  The
here obtained band  topologically coincides with the conduction band
result of Ref. \cite{matheiss}.
\begin{figure}[h]
\begin{center}
\includegraphics[width=1.6in,height=1.6in]{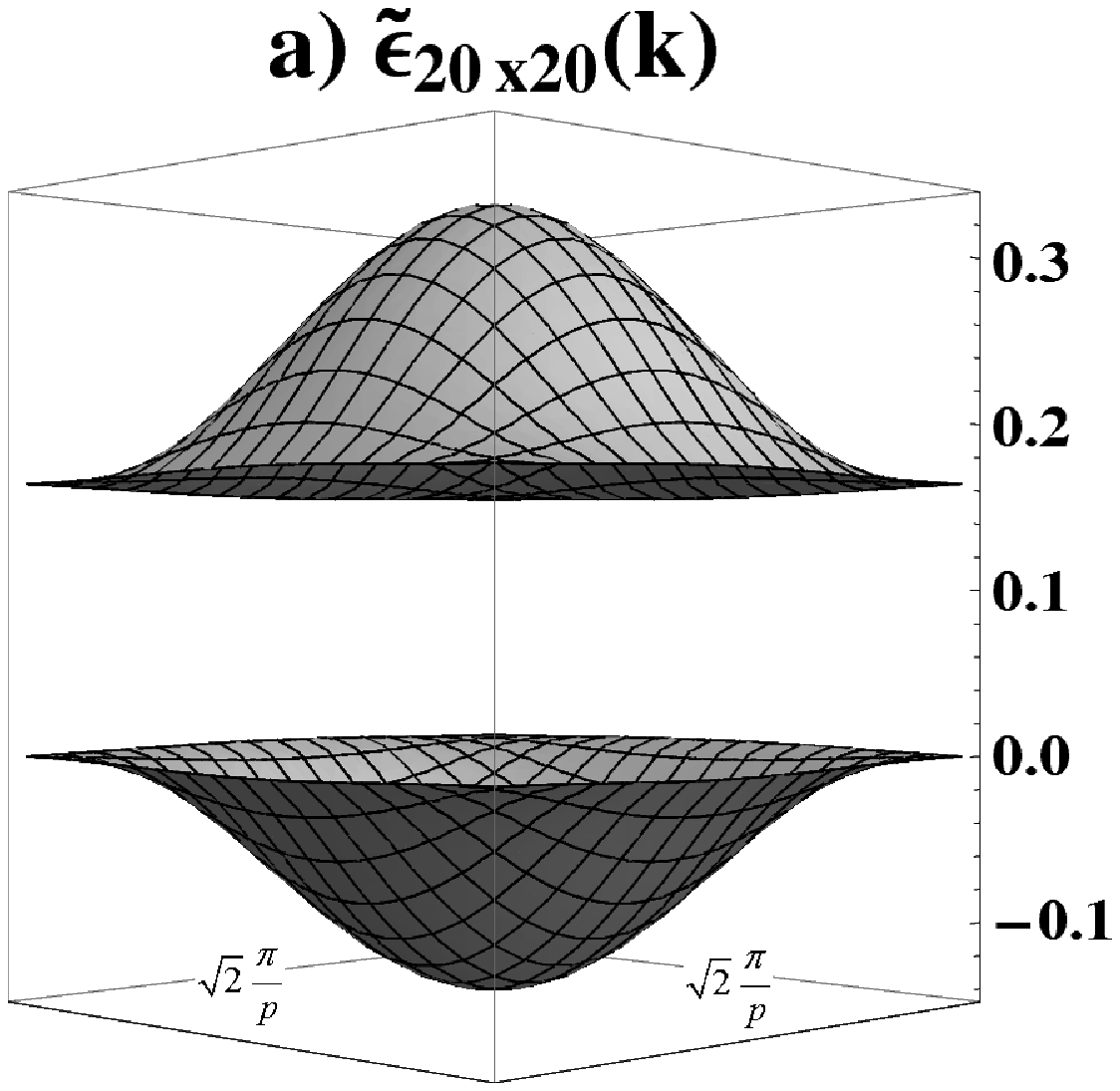}
\ \ \ \
\includegraphics[width=1.6in,height=1.6in]{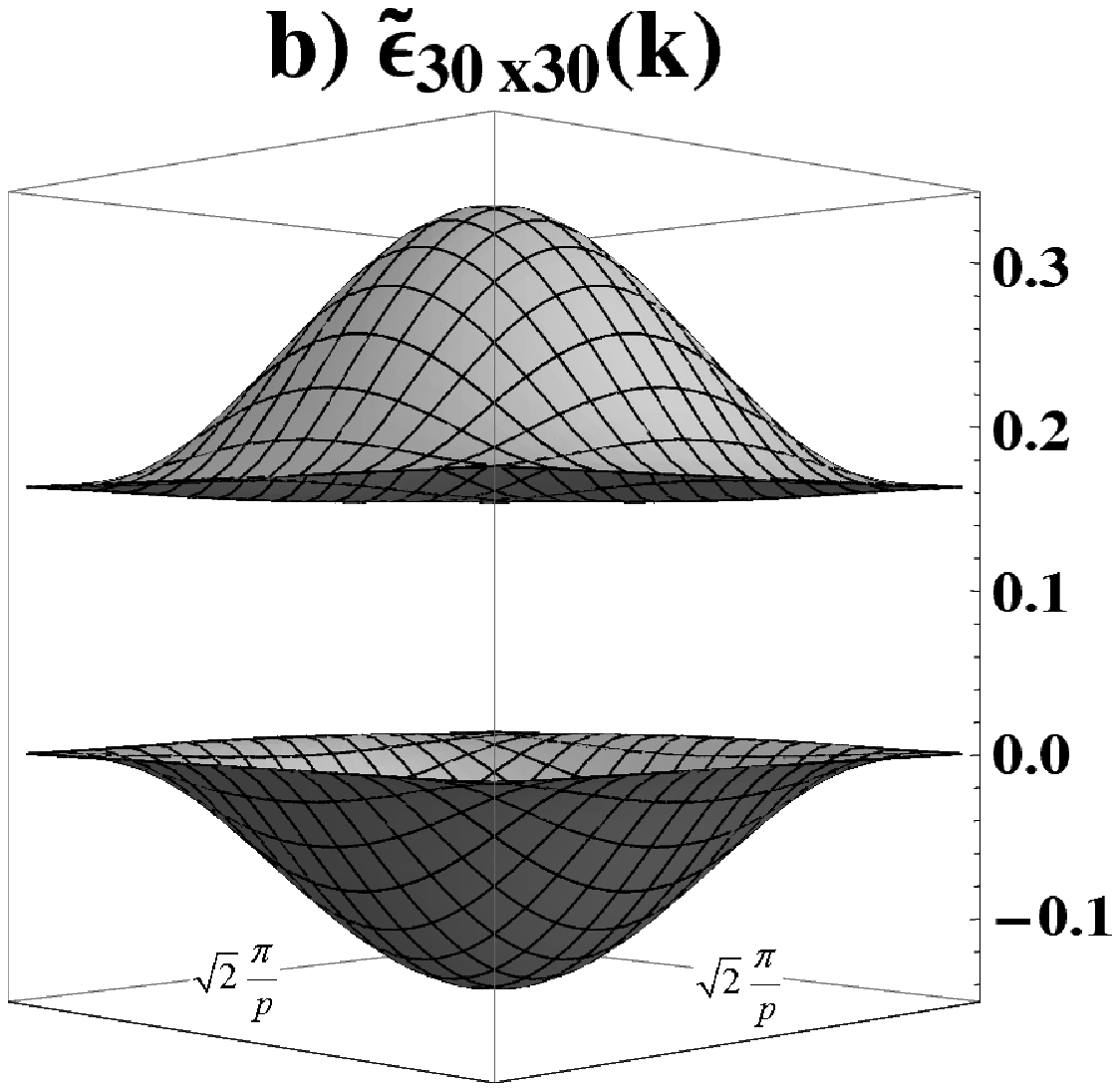}
\end{center} \caption{Energy bands obtained for: a) A sample of  20x20 cells,
 E$_{gap}$ = 1.32 eV. b) A sample of 30x30 cells, E$_{gap}$ = 1.32 eV.
 The zero energy level is chosen in the Fermi level of the
20$\times$20 system and the plotting region is the B.Z. of the
sublattices: the shadowed square of side $\sqrt{2}\pi/p$  in Fig. 2
a). } \label{f:bandas}
\end{figure} \\
\indent  After being determined the parameters of the model, we
searched for general HF solutions showing a non separable structure,
and moreover, also
  retaining a reduced translational symmetry in the defined sublattices.
The iterative process employed for finding this solution, started
from a particular state having an antiferromagnetic character from
the beginning (see Ref. \cite{sversion}). This initial form
helped the convergence toward the results being depicted in figure
\ref{f:bandas}. It shows two sets of bands following for a half
filling band, associated to  two periodicity lattices having
20x20 and 30x30 cells.  The difference between the energies in
them is of the order of 10$^{-5}$ dimensionless units of energy
$\frac{\hbar^{2}}{ma^{2}}$= 8.3 eV. The measure of the  gap shown
indicate that the bands correspond to insulating states. The close
similarity between both results indicates that the thermodynamical
limit has being achieved. The HF energy of this solution was the
lowest among of all the ones found. It also follows that in
coincidence with the experimental evidence, these states show a
local magnetic moment resting on the direction of the sublattice
x$_{12}$ as shown in  figure \ref{f:magnetizacion}. In the figure
\ref{f:magnetizacion} a) the only non vanishing component of the
magnetization $\textbf{m}$ in this solution is plotted. An
interesting result is that it has been experimentally observed
that La$_2$CuO$_4$ has a magnetic moment of 0.68 $\mu_B$ per  Cu
site on the CuO plane \cite{pickett}, and the value obtained from
evaluating the magnetization of this HF state turns out to be 0.67
$\mu_B$. Therefore, the considered here discussion satisfactorily
predicts the whole antiferromagnetic structure of La$_2$CuO$_4$.
The antiferromagnetic and isolator solution  will be referred
below as the IAF state.
\begin{figure}[h]
\begin{center}\label{f:magnetizacion}
\includegraphics[width=1.5in,height=1.5in]{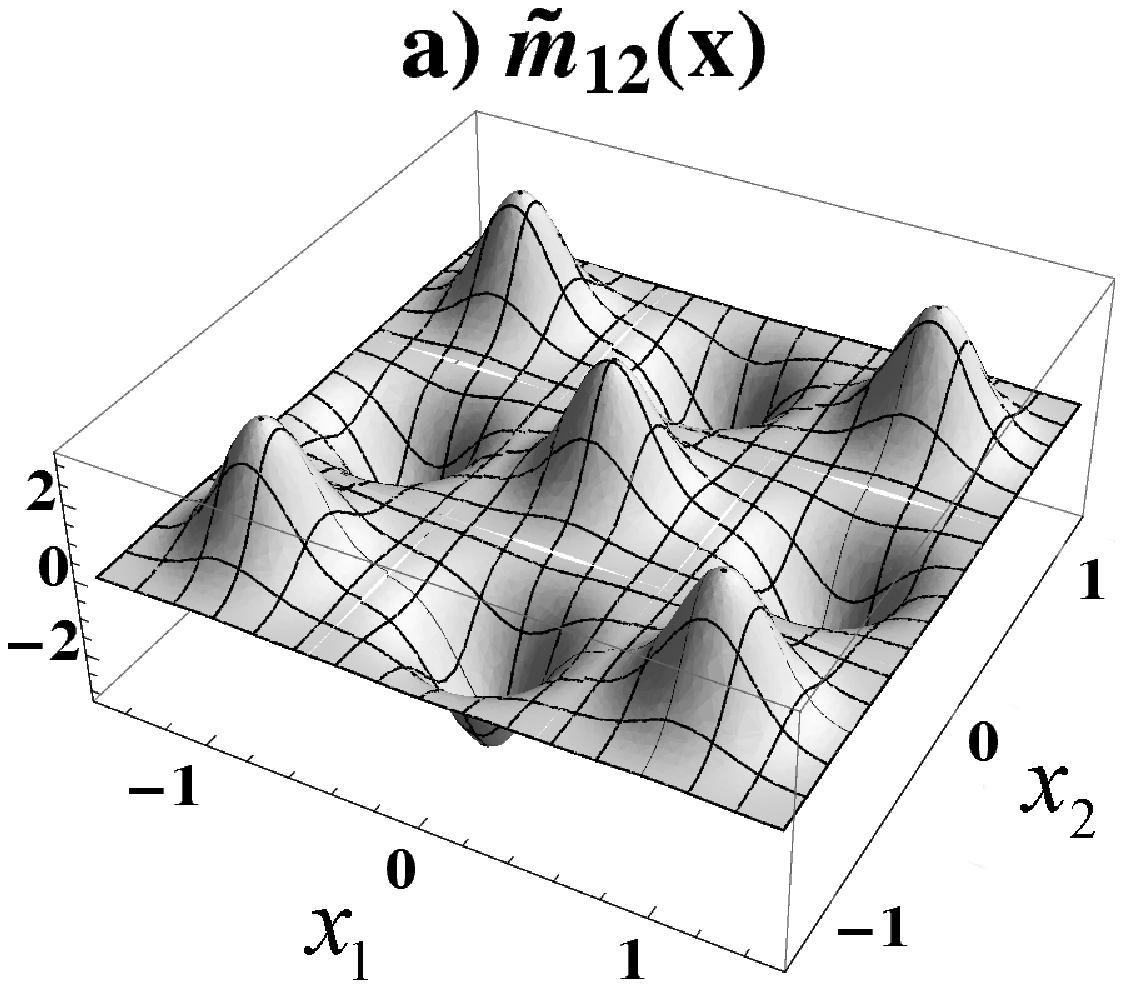}
\ \ \ \
\includegraphics[width=1.25in,height=1.25in]{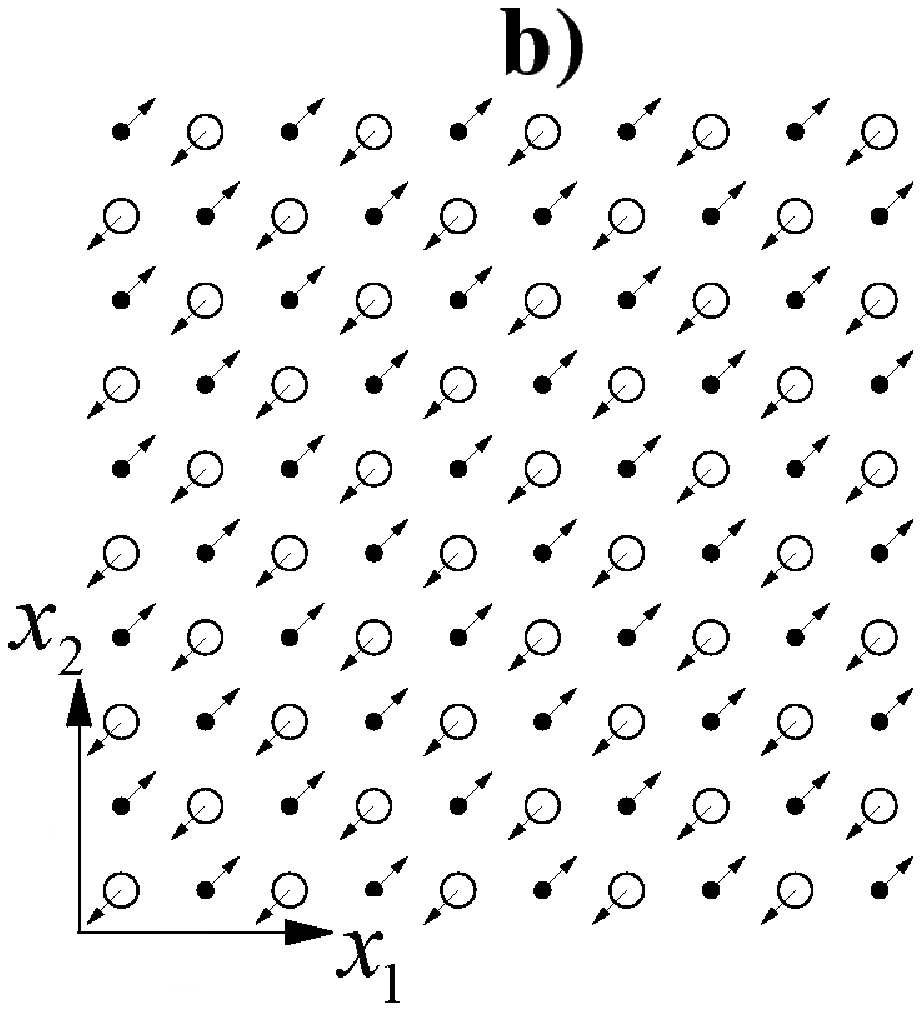}
\ \ \ \
\includegraphics[scale=0.25]{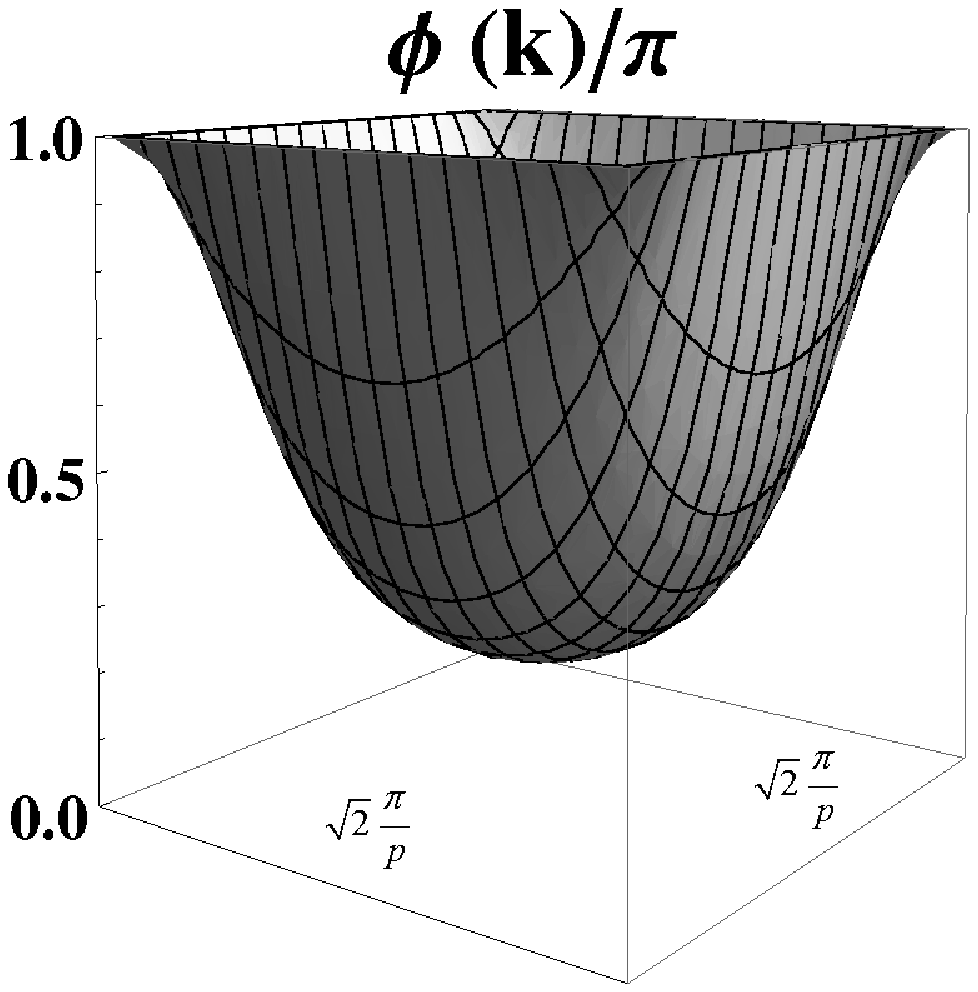}
\end{center} \caption{ The  magnetization vector  $\textbf{m}$
of the more stable HF state rests in the direction 1-2. a) The
figure shows the projection $\widetilde{m}_{12}$ of the
 dimensionless magnetization, in the  1-2 direction. The magnetization unit is
 $\frac{\mu_B}{p^2}$. b) The picture shows a scheme of the mean magnetic moment per site in the
 lattice. The figure insertion below depicts the angle between the magnetization components
 of each orbital on each of the sublattices. Note  that the orbitals are more AF like
 as they become close to the Fermi level. The plotting area is again the B.Z. of
 the sublattices shown in Fig. 2 a).}
\label{f:magnetizacion}%
\end{figure}
The single particle states of this solution carry a more intensive
antiferromagnetism as more closer they are in energy from the Fermi
surface. This property offers a clear  explanation of the gradual
loss observed in the antiferromagnetic order under the doping with
holes \cite{imada}. The dependence of the angle $\phi$ between the
magnetic moments per cell on each of the two sublattice shown by a
given Bloch orbital, is plotted in the insertion at the bottom of
figure \ref{f:magnetizacion}. These components are defined as the
integrals of the magnetic moment over all the unit cells of the
\textit{absolute} lattice centered in the sublattice points
\cite{sversion}. Then, this HF solution indicates that after the
orbitals are allowed to spatial dependent spin orientations, the
electrons prefer to reorient their spin when traveling between
contiguous lattice cells.
  Note that the states laying just on the Fermi surface are perfectly
antiferromagnetic ones, and that the more away from the boundaries
the orbitals are,  the less antiferromagnetic  they become.
\begin{figure}[h]
\begin{center}\label{f:bandas1}
\includegraphics[width=1.5in,height=1.5in]{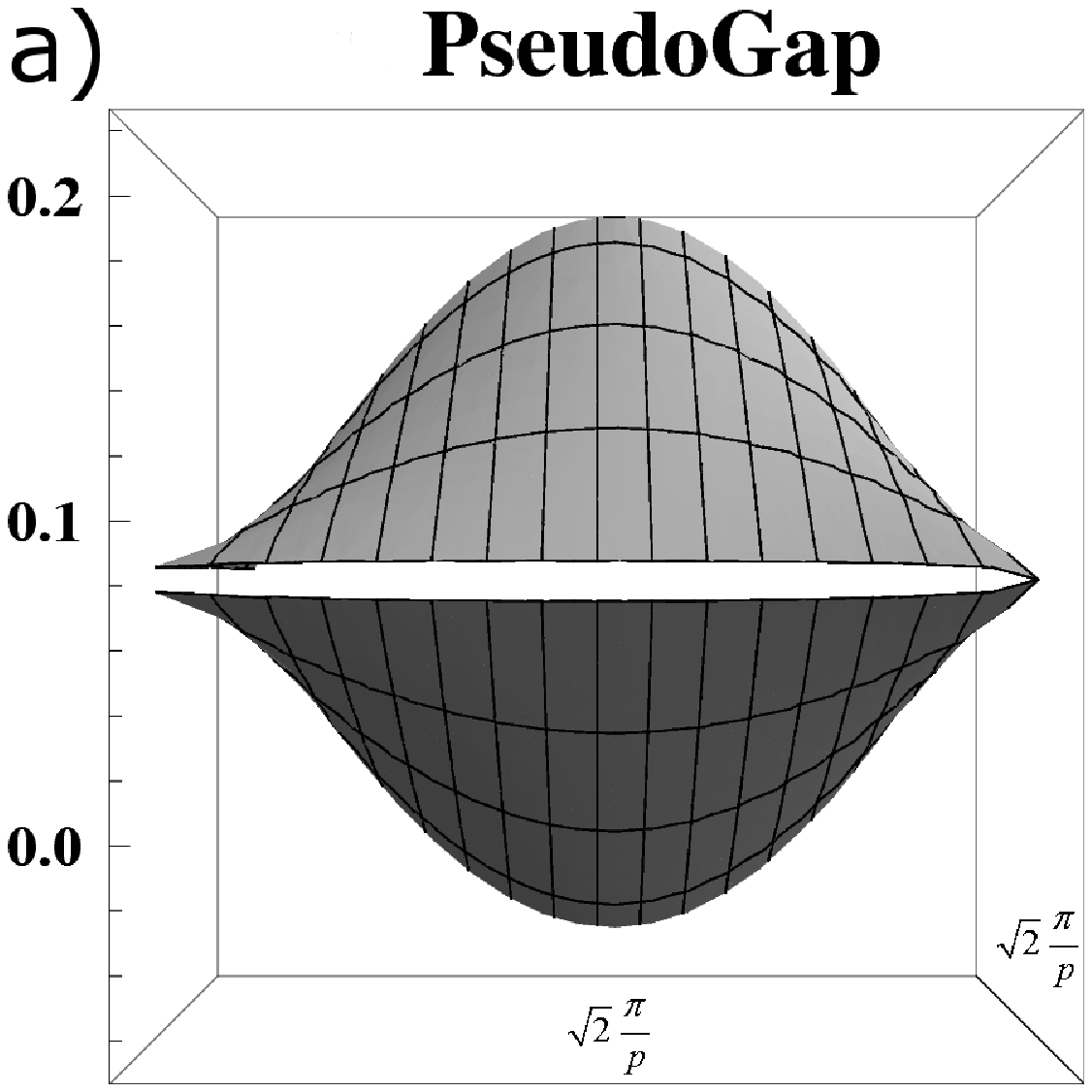}
\includegraphics[scale=0.35]{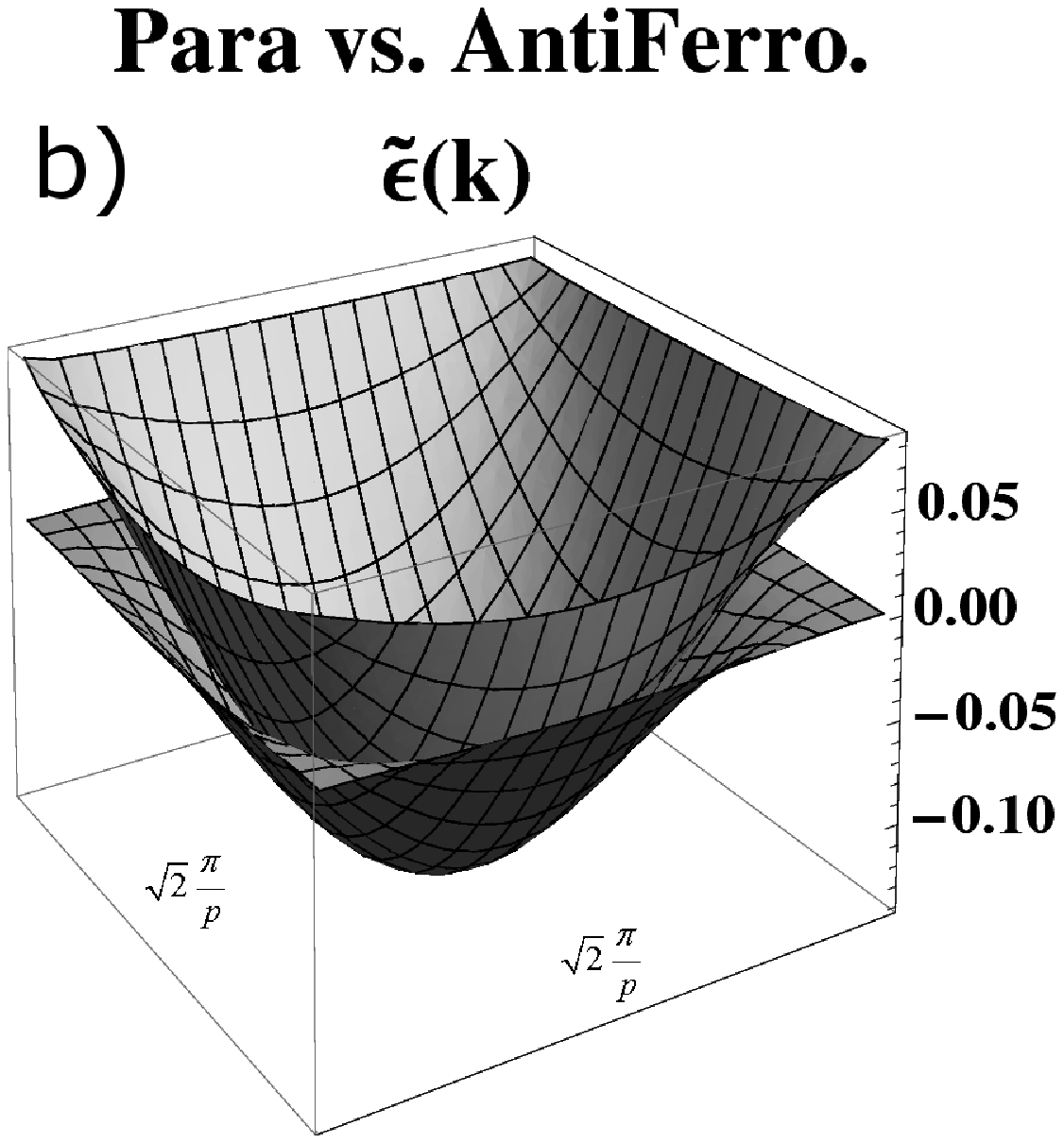}
\end{center} \caption{a)\ The band structure associated to
the paramagnetic ground state also showing a pseudogap. b) The
figure shows in the same plot the two occupied bands linked with the
pseudogap and the IAF states. The difference in their energies is
concentrated near the Fermi level.
 The zero energy reference coincides with the Fermi level of the IAF
 state in both pictures. The domains also in both cases is the B.Z.
 of the sublattices shown in Fig. 2 a).}
\label{f:bandas1}%
\end{figure}
\indent Further, in figure \ref{f:bandas1} a)  the band  spectra
corresponding to another  HF paramagnetic ground state (to be named
as the PPG one), obtained from the HF equations in Ref.
\cite{sversion} is shown. This solution was obtained by only
requiring the orbitals to show the full translational symmetry of
the CuO planes but not an $\alpha$ or $\beta$  type of their spin
structure. Note that simply allowing this last freedom determined
the existence of a pseudogap in the state. It reaches a maximum
value of 0.1 eV $\approx$ 10$^3$ K (equivalent to 0.012
dimensionless unit of energy $\frac{\hbar^2}{ma^2}$ = 8.3 eV). It is
an interesting outcome  that the  HF energy of this ground state is
exactly coincident with the one corresponding to the paramagnetic
and metallic state. Moreover, the occupied single particle states in
both solutions are identical and in consequence the momentum
dependence of the filled energy bands also coincide. Henceforth, the
difference between the two solutions only refers to the non occupied
states. It thus follow that the HF energies per particle of the
paramagnetic-metallic (PM), and the paramagnetic with pseudogap
(PPG) HF states coincide. They show a value +0.076l $eV$ higher than
the energy per particle corresponding to the
insulator-antiferromagnetic (IAF) ground state. The spin structure
of the excited states can be seen as a dynamical "correlation"
effect determined by the freedom allowed to the HF orbitals. It is
worth noticing
 that the energy difference PM (PPG)-IAF and the N\'eel temperature
of this kind of materials are both of the order of 10$^2$ K. Then,
the results suggest the possibility of having further success in
applying the approach started in this work to the description of
some regions of the phase diagram of the La$_2$ CuO$_4$. In Figure
\ref{f:bandas1} b)  the PPG (PM) and IAF occupied bands are depicted
in a common frame. The main difference in their energies corresponds
to the single particle states being closer to Fermi surface. As
noted before the same behavior has the antiferromagnetic character
of the single particle states of the IAF solution. Therefore, it  is
suggested that both solutions could evolve toward a common ground
state lacking an absolute magnetic order under under doping with
holes. This indication is in correspondence with the pattern shown
by the phase diagrams of these
materials \cite{pickett}.\\
 \indent  We express our gratitude to A. Gonz\'alez, C.
Rodr\'iguez, A. Delgado, Y. Vazquez-Ponce and N. G. Cabo-Bizet by
helpful conversations and comments. In addition, the relevant
support received from the Proyecto Nacional de Ciencias B\'asics
(PNCB, CITMA, Cuba) and from the Network N-35 of the Office of
External Activities (OEA) of the ASICTP (Italy) are deeply
acknowledged.

\end{document}